\documentclass[12pt]{article}
\usepackage{epsf}
  \newcommand{\be}{\begin{equation}}
\newcommand{\ee}{\end{equation}} \def\la{\mathrel{\mathpalette\fun
<}} \def\ga{\mathrel{\mathpalette\fun >}}
\def\fun#1#2{\lower3.6pt\vbox{\baselineskip0pt\lineskip.9pt
\ialign{$\mathsurround=0pt#1\hfil
##\hfil$\crcr#2\crcr\sim\crcr}}}

\newcommand{\bc} {\begin{center}}
\newcommand{\ec} {\end{center}}

\newcommand{\ver}{\mbox{\boldmath${\rm r}$}}

\newcommand{\veq}{\mbox{\boldmath${\rm q}$}}

\begin{document}

\bc
{\bf \LARGE Freezing of QCD coupling affects the short distance
static potential}
\ec
\vspace{.5cm}
\bc
{\large  A.M.Badalian\footnote{e-mail:
badalian@heron.itep.ru}, D.S.Kuzmenko\footnote{e-mail:
kuzmenko@heron.itep.ru} } \ec
\bc
{\large \it Institute of Theoretical and Experimental Physics,\\
 117218, B.Cheremushkinskaya 25, Moscow, Russia}
\ec

\begin{abstract}
A striking contradiction between lattice short range static potential
($n_f=0$) and standard perturbative potential, observed in Ref. [18], is
investigated in the framework of the background perturbation theory. With
the use of the background coupling $\tilde \alpha_B(r)$ which contains the
only background parameter - mass $m_B$, fixed by fine structure fit in
bottomonium, the lattice data are nicely explained without introduction of
exotic short range linear potential with large "string tension"
$\sigma^*\sim$ 1 GeV$^2$. A significant difference between $\tilde
\alpha_B(r)$ and standard perturbative coupling $\alpha_V(r)$ is
found in the range 0.05 fm $\la r\la 0.15$ fm, while at larger distances, $r
> 0.3$ fm $\tilde \alpha_B(r)$ fast approaches the freezing value $\tilde
\alpha_B(\infty)$. Some problems concerning the strong coupling properties
at short and long distances are discussed and solutions are suggested.
 \end{abstract}

\section{  Introduction}

The property of freezing of the strong coupling constant
$\tilde\alpha_V(r)$ at  long
distances is widely used in QCD phenomenology \cite{1}-\cite{6}. On the
fundamental level this phenomenon has been studied in two different
theoretical approaches \cite{7}-\cite{9}.  In the  case of the static
potential the freezing of the coupling $\tilde\alpha_V(r)$
suggests that $\tilde\alpha_V(r)$ is approaching a constant
$\alpha_{\mathrm{fr}}\equiv \tilde\alpha_V(r\to \infty)$ at relatively long
distances  while at small $r$ it manifests the property of asymptotic
freedom. Both characteristic features of the static potential were widely
used in hadron spectroscopy. However, it was realized that the asymptotic
freedom behavior does not practically affect hadron spectra being important
mostly for a wave function at the origin. On the contrary, the choice of
$\tilde\alpha_V(r)$
 as a constant at all distances, i.e.
  $\tilde\alpha_V(r)\cong \bar \alpha$, appears
  to be a reasonable approximation and gives rise to a good
description of meson spectra both for heavy quarkonia
\cite{1,10,11} and for heavy-light mesons \cite{12}.
 Also in  lattice QCD this choice gives
  a good fit to lattice static potential at the  distances above 0.2 fm.

Therefore the question arises why this simple approximation,
$\tilde\alpha_V(r)\approx \bar \alpha$,  works so well even in the
case of bottomonium where the  sizes of low-lying levels are not
large,   the characteristic radius   $R_{\mathrm{ch}}\la$
 0.5 fm.
  To answer this question one  needs to clarify another problem.
Namely, to find out  the  precise freezing value  of the vector
constant in momentum and coordinate spaces, and to define the
distances $r$ where  the difference between $\tilde\alpha_V(r)$ and
$\alpha_{\mathrm{fr}}$ is becoming inessential and therefore the
approximation $\tilde\alpha_V(r)=\bar \alpha
 ~(\bar\alpha\neq\alpha_{\mathrm{fr}}$ in general case) gives  a good
description of  hadron spectra and other physical characteristics.

This problem will be discussed in the present  paper in the framework of
background perturbation theory (BPTh) and it will be  shown that the
background coupling $\tilde\alpha_B(r)$ approaches its freezing value
already at rather small distances $r>0.4$ fm.  Here it is worthwhile
to remind that in most
 calculations in coordinate space the "average"
 value $\bar \alpha$  is usually taken in the range:
$0.35\la \bar \alpha\la 0.45$ while in momentum space  larger
critical values were used ($\alpha_{\mathrm{cr}}\equiv \alpha_V (q=0)$),
for example, in Ref. \cite{3} $\alpha_{\mathrm{cr}}=0.60$ and for the
Richardson potential $\alpha_{\mathrm{cr}}= \frac{2\pi}{\beta_0}\approx 0.7
~(n_f=3)$ \cite{2,13}, i.e. the  difference between
$\alpha_{\mathrm{cr}}\approx 0.6\div 0.7$ in momentum space and $\bar
\alpha\approx 0.40\pm0.05$  in coordinate space is essential.
However, this difference was not confirmed by the analysis of the
QCD coupling in background fields \cite{14} where the vector
 coupling constants  $\alpha_B(q)$ and $
\tilde\alpha_B(r)$ were found to  have the same asymptotic value:

\be
\alpha_B(q=0)=\tilde\alpha_B(r\to\infty) = \alpha_{\mathrm{fr}}.
\label{1}
\ee
 This equality takes place also for the phenomenological
coupling taken as a sum of Gaussians in Ref. \cite{3}.

In lattice measurements of the static potential at long  distances
 the freezing phenomenon is also seen, however, existing lattice data have not
clarified our knowledge about $\alpha_{\mathrm{fr}}$. As shown in Refs.
\cite{15}, \cite{16} lattice static potential at $r\geq 0.2$ fm
can be parameterized with a good accuracy by Cornell potential
with
 rather small $\bar \alpha$
(in lattice notation $\frac43\bar\alpha=e$).
 In quenched approximation $(n_f=0)$ the fitted  lattice values of
 $\bar \alpha
\approx 0.20\div 0.24 ~(e=0.27\div 0.32)$ turned out to be  small
so that in some cases there appears a discontinuity of the vector
coupling constant at the matching point, $r_{\mathrm{mat}}\approx 0.2$ fm
\cite{5}. But if dynamical fermions  are  introduced, in lattice QCD the
fitted value of $\bar \alpha$  $(n_f=2,3)$ was  found to become larger:
$ \bar \alpha\cong 0.30$ $(e\cong 0.40)$ \cite{17}  still being
less than in phenomenological models.

Another problem concerns behavior of $\tilde\alpha_V(r)$ at short distances.
The most  interesting and unexpected results were obtained  in
lattice measurements of the  static potential   at short
distances, 0.05 fm$\leq r\leq$ 0.15 fm \cite{18}, where a large
difference between lattice  and two-loop (one-loop) perturbative
potential was observed, yielding  discrepancy about 100\% at the
point $r=0.15$ fm.  In Ref.\cite{18} this large
 difference was parametrized by a short distance linear term   $\sigma^*r$
with the slope $\sigma^*\approx 1$ GeV$^2$.

This effect will be explained in our paper.
 To this end the
strong coupling constant in background fields $\tilde\alpha_B(r)$
will be calculated and the influence of the
background mass $m_B$ will be shown to become essential already at rather
short  distances. We have found that there is no need to introduce
an additional exotic linear potential $\sigma^* r$ as in Ref.[18]. Resulting
static potential $V_B(r)$ appears to have an effectively linear
term  in a good agreement with lattice data. In our calculations no fitting
parameters are introduced: the value of the background mass $m_B=1.0$ GeV is
taken from fine structure analysis in bottomonium \cite{11} while the QCD
constant $\Lambda_{\overline{MS}}(n_f=0)$ is considered as  a well
established number and taken from Ref.\cite{19}.

\section{The strong coupling constant $\alpha_B(q)$ in background field theory}

The perturbative static potential is used to define a  coupling constant
$\alpha_V(q)$ in $V$-scheme:
\be
V_P(q)=-4\pi C_F\frac{\alpha_V(q)}{q^2}, \label{2} \ee where
$q^2\equiv \veq^2$. Recently the renormalized $ \alpha_V(q)$ was
calculated in two-loop approximation \cite{20,21}. In coordinate
space the static potential  can be defined as the Fourier
transform of $V_P(q)$,
\be
V_P(r)= \int\frac{d\veq}{(2\pi)^3} V_P(q)\exp (i\veq\ver),
\label{3} \ee which gives rise to the simple  relation between the
coupling constants:

\be
\tilde\alpha_V(r)=\frac{2}{\pi}\int^\infty_0{dq}\frac{\sin
qr}{q}\alpha_V(q),
\label{4}
\ee
if
 the following
definition for  the coupling in coordinate space
$\tilde\alpha_V(r)$ is used,

\be
V_P(r)=-C_F\frac{\tilde\alpha_V(r)}{r}.
\label{5}
\ee

However, the Fourier transform of the perturbative  coupling
$\alpha_V(q)$, Eq.(\ref{4}), does not exist in a strict sence
because of Landau pole singularity. To escape IR divergency the
expansion of $\alpha_V(q)$ in the perturbative series at large
$q^2$ is usually made \cite{20,21}, but the resulting expansion is
valid only at short  distances.

Here we suggest to obtain the static potential in coordinate space
with the use of the coupling in
BPTh,  where  the vector coupling constant $\alpha_B(q)$ in
momentum space is defined at all momenta and has no singularity for $q^2>0$
\cite{7}. Then the potential in momentum space can be written as in
Eq.(\ref{2}),
 \be
V_B(q)=-C_F4\pi\frac{\alpha_B(q)}{q^2}.
\label{6}
\ee

In Eq. (6) and below  we consider the  influence of background
vacuum fields only on Coulomb-type interaction. In presence of
background fields the QCD coupling is modified so that it
 depends on the combination
 $(q^2+m^2_B)$ instead
  of $q^2$ as it is in standard perturbative theory \cite{8}.
  The mass $m_B$ is a background mass  which is
characteristic for a process considered.
    In two-loop approximation the running background coupling is
\be
\alpha_B^{(2)}(q)=\alpha^{(1)}_B(q) \left \{
1-\frac{\beta_1}{\beta^2_0}\frac{\ln t_B}{t_B}\right\},
\label{7}
\ee
where  one-loop  expression is given by
\be
\alpha_B^{(1)}(q)=\frac{4\pi}{\beta_0t_B},
 \label{8} \ee
  with
\be
t_B=\ln\frac{q^2+m^2_B}{\Lambda^2_V}.
\label{9}
\ee
 The condition $m_B>\Lambda_V$ is assumed to be satisfied under the
logarithm  (\ref{9}) to guarantee the absence of Landau pole; this
condition
 is always  valid for the numbers  $\Lambda_V$ and $m_B$ used in our
calculations  (see the numbers in Eq.(\ref{25a})).

In Eq.(\ref{7})
\be
\beta_0=11-\frac23 n_f; \beta_1=102-\frac{38}{3}n_f.
\label{10}
\ee

First  we discuss the  most important
 properties of the background coupling $\alpha_B(q)$.

 \underline{The background mass $m_B$} is not an arbitrary
 parameter. It can be calculated in the framework of BPTh or in
 lattice QCD. As was shown in Ref. \cite{22} the background mass
 $m_B$ in the case of the static potential is defined by  the
 difference  of two-gluon   and one-gluon hybrid excitations and
 can be extracted from the corresponding level  differences of
 hybrids $ c\bar c g, b\bar bg$. In Ref. \cite{22} this mass
 $m_B$ was evaluated to be 1.0$\div $1.2 GeV.
 For other processes, like $e^+e^-\to $ hadrons, in general case
 the background mass $m_B$ may be different \cite{8}.
 The appearence of the mass $m_B$ in Eq.(9) is similar to the case of QED
where $\alpha$ has the mass of $e^+e^-$ pair under logarithm.

 It is  of interest to notice that the analytical form of
 $\alpha_B(q)$ (\ref{7})  coincides with
 that  obtained in a picture when gluon is supposed  to have an effective
mass $m_g$ inside gluon loop. Therefore in Refs. \cite{23} $\alpha_s(q)$ was
taken as a function of ($q^2+4m^2_g$), i.e. the double effective gluon mass
$2m_g$ plays a role of the background mass $m_B$ (see a discussion
in Ref. \cite{4}).

In our calculations here the value of  $m_B$ will be fixed at $m_B=1.0$ GeV
taken from the fit to  fine structure splittings in bottomonium \cite{11}.

\underline{At large momenta, $q^2\gg m_B^2$,} the background
coupling goes over into the standard perturbative expression
 $\alpha_V(q)$. Therefore the QCD constants $\Lambda_{
 \overline{\mathrm{MS}}}$ in PQCD
and $\Lambda_{\overline{\mathrm{MS}}}^B$ in BPTh must coincide, in any case
it is true for the number of flavors $n_f=5$. As can be directly
calculated with the use of matching procedure  they are also equal
for $n_f=4$.

The QCD constant $\Lambda_V$, entering  the coupling
$\alpha_V(q)$ in Eq. (\ref{2}), can be expressed through
$\Lambda_{\overline{\mathrm{MS}}}(n_f)$ in $\overline{\mathrm{MS}}$
renormalization scheme \cite{24}:

\be
\Lambda^{(n_f)}_V=\Lambda^{(n_f)}_{\overline{\mathrm{MS}}}\exp\left(\frac{a_
1}{2\beta_0} \right)
\label{11},
\ee
with

\be
a_1=\frac{31}{3}-\frac{10}{9} n_f.
\label{12}
\ee
  At present the
values of $\Lambda^{(n_f)}_{\overline{\mathrm{MS}}}$ are well established
for the number of flavors $n_f=5$: $\Lambda_{
\overline{\mathrm{MS}}}^{(5)}=208\pm ^{25}_{23}$ MeV \cite{25} and  also
for $n_f=0$ due to analysis in lattice finite size technique
\cite{19}:
\be \Lambda^{(0)}_{\overline{\mathrm{MS}}}=\frac{602(48)}{r_0},
\label{13}
\ee
where $r_0 $ denotes the Sommer scale. With the
use of $r_0=2.5$ GeV$^{-1}$, taken  in most lattice
calculations \cite{17,18}, one obtains

\be
\Lambda^{(0)}_{\overline{\mathrm{MS}}}=241\pm 19 \mathrm{MeV},
\label{14}
\ee
then
from Eq. (\ref{11}) the QCD constant in the $V$-scheme is

\be
\Lambda^{(0)}_V=385\pm30 \mathrm{MeV}.
 \label{15}
 \ee

\underline{In three-loop approximation } the background coupling

\be
\alpha^{(3)}_B(q)= \alpha^{(1)}_B(q)\left\{1-
\frac{\beta_1}{\beta_0^2}\frac{\ln t_B}{t_B} +
\frac{\beta^2_1}{\beta^4_0t^2_B} \left[(\ln t_B)^2-\ln
t_B-1+\frac{\beta_2^V\beta_0}{\beta^2_1}\right]\right\},
\label{16}
\ee
contains the term including $\beta^V_2$
coefficient, which depends on a renormalization scheme and   was
calculated in Refs. \cite{20,21} (the coefficients
$\beta_0,\beta_1$ do not depend on the renormalization scheme); this
coefficient
 \be
\beta^V_2=\beta_2^{\overline{\mathrm{MS}}} -a_1\beta_1+(a_2-a_1^2)\beta_0.
\label{17}
\ee
Here $a_1$ is defined by Eq.(\ref{12}) and

\be
\beta^{\overline{\mathrm{MS}}}_2=\frac{2857}{2}-\frac{5033}{18}
n_f+\frac{325}{54} n_f^2,
 \label{18}
\ee
$$
a_2=9\left(
\frac{4343}{162}+4\pi^2-\frac{\pi^4}{4}+\frac{22}{3}\zeta(3)\right)+\frac{100}{81}n^2_f-
$$
\be
-\frac{3}{2}n_f\left( \frac{1798}{81}+\frac{56}{3}
\zeta(3)\right)-\frac{2}{3}n_f\left
(\frac{55}{3}-16\zeta(3)\right)
 \label{19}
\ee
In Eq.(\ref{19})
$\zeta(3)=1.202057$  denotes the Riemann $\zeta$-function. In
quenched approximation

\be
a_2(n_f=0) =456,7488,~~~ \beta_2^V(n_f=0)= 4224,1817,
 \label{20a}
\ee
i.e. $\beta^V_2$ coefficient turns out to be  about 3 times larger than
$\beta_2^{\overline{\mathrm{MS}}} (n_f=0)=\frac{2857}{2}$. As a result
     the third order correction in   $\alpha_B^{(3)}(q)$ is much larger
than  $\alpha_B^{(1)}(q)$ and  $\alpha_B^{(2)}(q)$.

It is easy to find the first order correction to the perturbative
 coupling $\alpha_V(q)$ which
comes from the expansion of $\alpha_B(q)$ in  powers of
$m^2_B/q^2$. In two-loop approximation
 \be
 \alpha_{\mathrm{appr}}^{(2)}({\rm large}~ q)
=\alpha_V^{(2)}(q)
 -\alpha_V^{(1)}(q)\frac{m_B^2}{q^2\ln
\frac{q^2}{\Lambda^2_V}};
 \label{21a}
 \ee
 with
\be
\alpha_V^{(2)}(q)=\alpha_V^{(1)}(q)
(1-\frac{\beta_1}{\beta_0^2}\frac{\ln y}{y}),~~
\alpha_V^{(1)}(q)=\frac{4\pi}{\beta_0 y},~~
y=\ln\frac{q^2}{\Lambda^2_V}.
 \label{22a} \ee
This approximation appears to be valid only at $q>2$ GeV (with the accuracy
$\la$ 10\%).

\underline{The behavior of $\alpha_B(q)$ in IR region}. The
freezing value of $\alpha_B^{(n)}(q)$ can be easily obtained from
Eqs. (\ref{7})-(\ref{9}), in particular, for two-loop coupling
\be
\alpha^{(2)}_{\mathrm{cr}}=\alpha_B^{(2)}(q^2=0)=\frac{4\pi}{\beta_0t_0}
\left\{ 1-\frac{\beta_1}{\beta_0^2}\frac{\ln t_0}{t_0}\right\} \label{23a}
\ee  with
\be
\alpha^{(1)}_{\mathrm{cr}}=\frac{4\pi}{\beta_0t_0};~~t_0=t_B(q^2=0)= \ln
\frac{m^2_B}{\Lambda_V^2}.
\label{24a}
\ee
In what follows  the
notation $\alpha^{(n)}_{\mathrm{cr}} =\alpha_B^{(n)}(q^2=0)$, as in
potential models \cite{3}, is also used.
 The parameters $m_B$ and $\Lambda_V$,  present in $t_B$, are considered to
 be
 fixed:
the value of $m_B$ is taken from fine structure analysis of $2P$-
 and $1P$- states in bottomonium while the value
 $\Lambda_V^{(0)}(n_f=0)$
  is taken from lattice data and   given by  Eq.(\ref{15}),
   \be
 m_B=1.0 ~\mathrm{GeV},~~ \Lambda_V^{(0)}(n_f=0)=385~ \mathrm{MeV}.
 \label{25a}
 \ee
We suppose here that for $n_f=3$ the constants $\Lambda^{(3)}_V$
and $\Lambda^{(0)}_V$ are  approximately equal,
\be
\Lambda_V(n_f=3)\simeq\Lambda_V(n_f=0)=385\pm \mathrm{30 ~MeV}
\label{26a}
\ee
Then the  following critical values can be obtained from Eqs.
(\ref{7}),(\ref{15}):
\be
\alpha_{\mathrm{cr}}^{(1)}=0.598;~~ \alpha_{\mathrm{cr}}^{(2)}=0.428;~~
\alpha_{\mathrm{cr}}^{(3)}=0.805~~ (n_f=0)
\label{27a}
\ee

\be
\alpha_{\mathrm{cr}}^{(1)}=0.731;~~ \alpha_{\mathrm{cr}}^{(2)}=0.536;~~
\alpha_{\mathrm{cr}}^{(3)}=0.972~~ (n_f=3)
 \label{28a}
\ee

As one can see from Eqs.(\ref{27a}),(\ref{28a}) the third order coupling
turns out to be about 90\% $(n_f=0)$ and 80\% $(n_f=3)$ larger than
$\alpha_{\mathrm{cr}}^{(2)}(n_f)$ because of large $\beta^V_2$ coefficient
(20). Such a value of $\alpha^{(3)}_{\mathrm{cr}}(n_f)$ appears to be
too large and not compatible with the effective Coulomb constant of the
static potential, $\alpha_\mathrm{eff}\la 0.5$, needed to describe heavy
quarkonia \cite{1,3} and heavy-light meson spectra \cite{12}. Such large
value of Coulomb constant is also not observed in lattice calculations of
the static potential at large distances [15,16].

Presumably it means that since the
perturbative series is an asymptotic one, it should be truncated after
the second-loop term.

Recently an "analytical" perturbation theory was elaborated in papers [9].
It was shown there that the modification of singularities of perturbative
coupling by power terms allows to work accurately in two-loop
approximation.  The third order contribution becomes numerically
inessential and e.g. for $e^+e^-$-annihilation is about 0.5\%.

Therefore in what follows the third order term will be
omitted and  {\it a posteriori} our phenomenological analysis with the use
of background coupling will demonstrate that the two-loop approximation is
sufficient to describe the lattice data at small distances.

The background coupling $\alpha_B^{(2)}(q)$ in
two-loop approximation turns out to be rather close to the
phenomenological $\alpha_{\mathrm{ph}}(q)$ which is successfully used in
hadron spectroscopy. For comparison $\alpha_{\mathrm{ph}}(q)$  will be
taken from the well-known paper of Godfrey and Isgur \cite{3}:
\be
\alpha_{\mathrm{GI}}=0.25\exp(-q^2)+0.15\exp (-0.1q^2)+0.20\exp (-0.001
q^2)
 \label{30a}
 \ee
with $q$ in GeV and $ \alpha_{\mathrm{cr}}=0.60.$

In Fig.1 this phenomenological coupling is compared to the
background coupling $\alpha_B^{(2)}(q)$ for $n_f=3$. Here, as in
Eq.(\ref{26a}), it is supposed that
$\Lambda_V(n_f=3)\cong\Lambda_V(n_f=0)$ and for $\Lambda_V(n_f=3)$
two values, are taken:
$$
(A)~~\Lambda_V(n_f=3)=385~ \mathrm{MeV}
$$
\be
(B)~~\Lambda_V(n_f=3)=410~ \mathrm{MeV}.
 \label{31a}
\ee
These values of $\Lambda_V^{(3)}$ do not contradict those which are commonly
used  in
$\mathrm{\overline{MS}}$ renormalization scheme, and give rise to
$\alpha_s(M_Z)=0.118\pm 0.001$. The connection between
$\Lambda_{\mathrm{\overline{MS}}}^{(n_f)}$ and $\Lambda_V^{(n_f)}$ is given
by Eq.(11).

In Fig.1 the couplings
$\alpha_{\mathrm{GI}}(q)$ (solid line),
$\alpha^{(2)}_B(q)$  in two-loop approximation with $\Lambda_V=385 $ MeV
 (dashed line), and $\alpha_B^{(2)}(q)$ with $\Lambda_V=410 $ MeV
(dash-dotted line) and $m_B=1.0 $ GeV are shown. At the momentum $q=\bar
m_c=1.3$ GeV $(\bar m_c$ is the running mass of $c$ quark) the matching of
the couplings was done with the  following result for  the QCD constant:
$\Lambda_V(n_f=4)=0.325$ MeV in the case $A$ and $\Lambda_V(n_f= 4)=351 $
MeV in the case  $B$. As is  seen  from Fig.1 the background coupling
$\alpha_B^{(2)}(q)$ in the case $B$ (and to some extent in the
case A) appears to be very close to the phenomenological coupling
$\alpha_{\mathrm{GI}}(q)$; the difference between them is  less than 5\% at
small $q\la 1.3$ GeV and  less than 2\% in the range
 $1.3\leq q\leq 4 $ GeV. So
 one can expect that with the use
  of the background coupling as good
   description of low-energy experimental data can be obtained
as in Ref. \cite{3} with the use
  of the phenomenological coupling. From here our estimate of the freezing
value is about 0.53$\div$ 0.60, and it is interesting to look at the
expansion of the background coupling $\alpha_B(q)$ near the freezing point
$q=0$:
\be
\alpha_{\mathrm{appr}}^{(2)} ({\rm small}~ q) =\alpha_{\mathrm{cr}}^{(2)}
-\alpha_{\mathrm{cr}}^{(1)} \frac{q^2}{m^2_B\ln \frac{m^2_B}{\Lambda_V^2}},
\label{32a}
\ee
where $\alpha^{(2)}_{\mathrm{cr}}, \alpha_{\mathrm{cr}}^{(1)}$ are
the fixed numbers defined by $m_B$ and  $\Lambda_V$
(see Eqs. (\ref{23a}), (\ref{24a})). This approximation appears to
be valid only in very narrow range of small momenta $q$, $0\leq
q\leq 0.4$ GeV, where the difference between
$\alpha_B^{(2)}(q)$ and $\alpha_{\mathrm{appr}}^{(2)}({\rm small}~q)$ is
less than 5\%; but it already reaches 22\% at $q=1.0$ GeV.

\underline{Heavy-quark initiated jets} can be successfully
described at small momenta if the following assumption is made
about an effective coupling constant $\alpha_{\mathrm{eff}}(q)$ in infrared
region \cite{6}:

\be
J_2{(\mathrm{fit})}=(2~\mathrm{GeV})^{-1} \int^{2~\mathrm{GeV}}_0 dq
\frac{\alpha_{\mathrm{eff}}(q)}{\pi} =0.18\pm 0.01 (\exp) \pm
0.02(\mathrm{th}).
\label{33a}
\ee
As was shown in Ref.\cite{6} this number
does not depend on the form of $\alpha_{\mathrm{eff}}(q)$ assumed in the
fit. Our calculations of the integral (\ref{33a}) with the background
coupling $\alpha_B^{(2)}(q)$ from  Eq.(\ref{7}) with $n_f=3,\Lambda_V=410$
MeV, $m_B=1.0$ GeV  give

\be
J_2(\alpha_B)=(2~\mathrm{GeV})^{-1}\int^{2~\mathrm{GeV}}_0\frac{dq}{\pi}
\alpha_B^{(2)}(q)=0.134
 \label{34a}
 \ee
 i.e. this number is by
26\% smaller than $J_2$(fit) in  Eq.(\ref{33a}). The same
integral calculated with  the phenomenological constant
$\alpha_{\mathrm{GI}}(q)$ (\ref{30a})
 is also by   22\%
smaller than $J_2$(fit) (the central value):

 \be
 J_2(\alpha_{\mathrm{GI}})=0.14
\label{35a}
 \ee
  and very close to our
number (\ref{34a}). One should  notice here that the large number
(\ref{33a}) in Ref.[6] could be connected with the large  fitted  value
of  $\alpha_s^{\overline{\mathrm{MS}}}
(M_Z)=0.125\pm 0.003(\exp)= 0.004 (\mathrm{th})$, used in their paper,
while  now the average $\alpha_s^{ \overline{\mathrm{MS}}}(M_Z)=0.118\pm
0.001$ is accepted \cite{25}.

In conclusion we give our predictions  about the freezing values
 in momentum space:
$$
\alpha_{\mathrm{cr}}^{(1)}=0.598,~~\alpha_{\mathrm{cr}}^{(2)}=0.428~~
(n_f=0,~~\Lambda_V^{(0)}=385~\mathrm{MeV}),$$ $$
\alpha_{\mathrm{cr}}^{(1)}=0.731,~~\alpha_{\mathrm{cr}}^{(2)}=0.536~~
(n_f=3,~~\Lambda_V^{(3)}=385~\mathrm{MeV}),$$ \be
\alpha_{\mathrm{cr}}^{(1)}=0.783,~~\alpha_{\mathrm{cr}}^{(2)}=0.582~~
(n_f=3,~~\Lambda_V^{(0)}=410~\mathrm{MeV})
\label{36a}
\ee

\section{The background coupling $\tilde \alpha_B(r)$ in coordinate
space}

From the explicit   expressions of $\alpha_B^{(n)}(q)$ it is
evident that the coupling $\alpha_B^{(n)}(q)$ is well
defined at all momenta $q^2>0$ if the condition $m_B>\Lambda_V$ is
satisfied.
 Therefore the  Fourier transform can be used to define the
static potential in coordinate space over all distances:
\be
V_B(r)\equiv
-C_F\frac{\tilde\alpha_B(r)}{r}=-C_F4\pi\int\frac{\alpha_B(q)}{q^2}
e^{i\veq\ver}\frac{d\veq}{(2\pi)^3}
 \label{22}
 \ee
From here the relation
similar to Eq.(\ref{4}) follows:
\be
\tilde \alpha_B(r)=\frac{2}{\pi}\int^\infty_0 dq\frac{\sin qr}{q}
\alpha_B(q)=\frac{2}{\pi}\int^\infty_0 dx\frac{\sin
x}{x}\alpha_B(x/r),
 \label{23}
\ee
 where now  the background
coupling $\alpha_B(x)$ depends on the variable
\be
t_B(x)=\mathrm{ln} \frac{x^2+m^2_Br^2}{\Lambda^2_Vr^2} \label{24} \ee

This  integral (\ref{23}) cannot be taken analytically even in
one-loop approximation and was calculated numerically
 in $n$-loop approximations ($n=1,2$) with the use of
the parameters (\ref{25a}). The behavior of  background coupling
$\tilde \alpha_B^{(2)}(r)$ and $\tilde \alpha_B^{(1)}(r)$ is shown
in Fig.2 in the range $0\leq r \leq 1.4$ fm.

From Fig.2 one can see that  two-loop   background coupling in
coordinate space is approaching  the freezing value at relatively
short distances, $r\ga 0.4$ fm, and the values of
$\tilde\alpha_B^{(n)}(r)$ $(n=1,2)$ at the Sommer scale $r_0
\approx 0.5$ fm are following,

\be
\tilde \alpha_B^{(1)}(r_0=0.5~
\mathrm{fm})=0,574;~~\tilde\alpha^{(2)}_B(r_0=0.5~\mathrm{fm}) =0.404
\label{26}
 \ee
 It is of interest to notice that  the two-loop coupling at
the distance $r_0$  practically coincides with the number $\bar
\alpha=0.39$ widely  used in Cornell potential \cite{1}, while the
one-loop coupling is too large. This fact that two-loop background
coupling is almost constant already at $r\ga0.4$ fm can be considered as an
important argument in favour of the choice $\tilde\alpha_B(r)=$const in low
energy spectroscopy. In Fig.3 $\tilde
\alpha^{(2)}_B(r)$ is shown for two different values of the
background mass: $m_B=1.0$ GeV (solid line) and $m_B=1.1$ GeV
(dashed line). As seen from Fig.3 the difference between them is
becoming essential already at  $r\approx 0.3$ fm, being about 10\%
over all distances $r>0.3$ fm; their freezing values are
$\alpha_{\mathrm{fr}} ^{(2)}(m_B=1.0$ GeV)=0.428,
$\alpha_{\mathrm{fr}}^{(2)}(m_B=1.1$ GeV)=0.382 (in both cases
$\Lambda_V^{(0)}=385$ MeV).

 The freezing value in coordinate space
 turns out to be just the same as in momentum space, i.e. for
$n_f=3$ and $n_f=0$ $(\Lambda_V=385$ MeV)
 they are given by the numbers from Eq. (\ref{36a}).
This property is true also   for the  phenomenological potential
used in Ref.\cite{3}:
\be
\alpha_{\mathrm{GI}}(q=0)=\alpha_{\mathrm{GI}}(r\to\infty)=0.60,
 \label{28}
\ee
since  in coordinate space the coupling $\alpha_{\mathrm{GI}}(r)$
corresponding to $\alpha_{\mathrm{GI}}(q)$ in Eq. (\ref{30a}) is
\be
\alpha_{\mathrm{GI}}(r)=0.25 \Phi(2r)+ 0.15 \Phi(1.581 r) +0.20
\Phi(15.811r),
  \label{29a}
  \ee
  where $\Phi(z)$ is the error function. Thus
 for three flavors  the
phenomenological value $\alpha_{\mathrm{fr}}\approx 0.6$ was found to be a
bit  larger  than   our number $\alpha_{\mathrm{cr}}^{(2)}=0.54$ (see Eq.
(\ref{36a}) for $n_f=3$).

With $\tilde \alpha_B(r)$ calculated  above we can compare the
background potential $V_B(r)$ (37) to  the lattice static
potential  from  Ref.\cite{18}. Here we are mostly  interested in
short range  potential, in particular, in the  influence
 of background mass $m_B$ on its behavior.
The properties of $\tilde \alpha_B(r)$ at small $r$ will be
considered in the next  Section.

\section{$\tilde\alpha_B(r)$ at short distances}

Recently very precise lattice measurements of the static potential
at short distances were presented \cite{18}. Having   these data
one has a unique opportunity to compare theoretical
 predictions about  the background coupling and the potential $V_B(r)$  with
precise lattice data in quenched approximation.
 We remind that our calculations of  $\tilde\alpha^{(n)}_B(r)$
  $(n=1,2) $ were done without any arbitrary parameter:
$\Lambda_V^{(0)} =385$ MeV $(n_f=0)$  was fixed from lattice data
(Eq.(\ref{14})) and $m_B=1.0$ GeV from the fine structure splitting of $1P$
and $2P$ states in bottomonium.  In   Fig.4  $\tilde\alpha^{(n)}_B(r)$
is compared to the perturbative running coupling $\tilde\alpha_V^{(2)}(r)$
at $r\leq 0.12$ fm, $\tilde\alpha_V^{(2)}(r)$  is given by the
expression ($r\Lambda_R\ll 1$)

\be
\tilde\alpha_V^{(2)}(r)=\tilde\alpha^{(1)}_V(r)
\left\{1-\frac{\beta_1}{\beta_0^2} \frac{\ln y_R}{y_R}\right\}
\label{4.1} \ee
with
\be
\tilde\alpha^{(1)}_V(r)=\frac{4\pi}{\beta_0y_R},~~~
y_R=\ln\frac{1}{\Lambda_R^2r^2},
 \label{4.2}
\ee
and the following
prescription for the value of  QCD constant $\Lambda_R{(n_f)}$
\cite{24},\cite{13}
\be
\Lambda_R^{(n_f)}=\Lambda_V^{(n_f)}\exp \gamma_E \label{4.3} \ee
In Eq. (\ref{4.3}) $\gamma_E$ is the Euler constant
($\gamma_E=0.5772157$), and in quenched approximation $\Lambda_V^{(0)}$
is  given by the number (\ref{14}), therefore for
$n_f=0$
\be
\Lambda^{(0)}_R=684\pm53~\mathrm{MeV}
 \label{4.4}
\ee
In our calculations
below we take  the number

\be
\Lambda_R^{(0)}=686~\mathrm{MeV},
\label{4.5}
\ee
which
 corresponds to $\Lambda_V^{(0)} =385$ MeV according to the
relation (\ref{4.3}).

The numerical  comparison of the "exact" background coupling
$\tilde \alpha_B^{(2)}(r)$ ($\Lambda_V=385$ MeV, $n_f=0$) and the
corresponding perturbative $\tilde\alpha_V^{(2)}(r)$ with
$\Lambda_R^{(0)}$ from Eq. (\ref{4.5}) is presented in Table
for the distances in the range
\be
0.002~\mathrm{fm}\leq r\leq 0.15~\mathrm{fm}.
\label{4.6}
\ee

\vspace{1cm}

{\bf Table } \underline{The background coupling
$\tilde\alpha_B^{(2)}(r)$ ($\Lambda_V^{(0)}=385$ MeV,}\\
\underline{$m_B=1.0$ GeV) compared to perturbative coupling
$\tilde\alpha_V^{(2)}(r)$ with $\Lambda^{(0)}_R=686$ MeV.}

\vspace{1cm} \begin{center}

\begin{tabular}{|l|l|l|l|l|l|}\hline
$r$ in fm& $\tilde\alpha^{(2)}_B(r)$& $\tilde\alpha^{(2)}_V(r)$& $r$ in
fm& $\tilde\alpha^{(2)}_B(r)$& $\tilde\alpha^{(2)}_V(r)$\\\hline
0.002&0.0964&0.0924&0.024&0.1757&0.1667\\ \hline
0.004&0.11095&0.10505&0.030&0.1880&0.1807\\ \hline
0.006&0.1216&0.1209&0.0355&0.1988&0.1943\\ \hline
0.008&0.1303&0.1221&0.041&0.2085&0.2079\\\hline
0.012&0.1446&0.1352&0.049&0.2204&0.2264\\\hline
0.016&0.1564&0.1465&0.057&0.2311&0.2455\\\hline
0.020&0.1666&0.1569&0.063&0.2384&0.2605\\\hline
\end{tabular}
\end{center}
\vspace{1cm}

 One can see that the difference between these two couplings,
\be
\Delta \alpha^{(2)}_B(r)=\tilde
\alpha^{(2)}_B(r)-\tilde\alpha_V^{(2)}(r), ~~(n=1,2),
\label{50}
\ee
 has several prominent
features.

\underline{First, at very short distances}, $r<0.04$ fm, the
correction $\Delta\alpha^{(2)}_B(r)$ is \underline{positive, }
i.e.
\be
\tilde \alpha_B^{(2)}(r)>\tilde\alpha^{(2)}_V(r)~~(r<0.04~ \mathrm{fm}),
 \label{51}
\ee
 and relatively small. It is about 6\% at $r=0.02$ fm and still
remains $\sim 4$\% at much smaller $r=0.002$ fm so that
$\tilde\alpha^{(2)}_B(r)$ approaches the perturbative running coupling
$\tilde\alpha_V^{(2)}(r)$ rather slowly.
In one-loop approximation this correction was calculated
analytically  in Ref.[14]:
 \be
\Delta
\alpha^{(1)}_B(r)=\frac{\pi^3}{6\beta_0[-\ln(\Lambda_Vr)]^3}, ~~
|\ln(\Lambda_Vr)|\gg 1.
 \label{51a}
 \ee
It is positive
and less than 5\% only at very short distances, $r<0.007$ fm. The important
feature of  $\Delta\alpha^{(1)}_B(r)$ is that it does not
depend on the background mass $m_B$ in limit $r\to 0$.

\underline{Secondly, the values of $\tilde\alpha^{(2)}_B(r)$ and
$\tilde\alpha^{(2)}_V(r)$ coincide} at the point $r=0.041$ fm, i.e.
\be
\Delta \alpha^{(2)}_B(r=0.041~\mathrm{fm})=0
\label{52}
 \ee
At bigger distances, in particular, in the range
\be
0.04~\mathrm{fm} < r \la 0.15~\mathrm{fm}
 \label{53}
 \ee
 \underline{this
correction is negative and fast growing,} e.g  it is 13\% at $r=0.07$ fm,
already 36\% at  $r=0.10 $ fm and  reaches 100\% at $r=0.14$ fm
although all these points lie rather far from the Landau pole,
$r_{\mathrm{pole}}= 0.29$ fm.

The explanation why the perturbative coupling   is essentially larger than
the background coupling at rather small  $r$ was given in Ref.\cite{14}. It
was shown there that in coordinate space the  QCD constant $\Lambda_R$ can
be defined  as a constant (given by Eq.(45)) only at very short
distances while in the  transition region (\ref{53}) the role of
QCD "constant" plays a function  $\tilde \Lambda_R(r)$ dependent
on the distance:
\be
\Lambda_R\to \tilde \Lambda_R(r)\cong \Lambda_V\exp
(\gamma_E+\sum^\infty_{k=1}\frac{(-m_Br)^k}{k!k})\cong
\Lambda_R\exp (-m_Br+\frac{1}{4} m^2_Br^2) \label{54}
\ee
Then
with the use of the function $\tilde \Lambda_R(r)$ the perturbative
coupling $\tilde\alpha^{(2)}_V(r)$  reproduces $\tilde \alpha_B^{(2)}(r)$ in
the range (\ref{53}) with the accuracy better than 5\%. Actually, this
approximation (\ref{54}) can be used only at  distances
\be
m_Br-\frac{m^2_Br^2}{4}<\gamma_E,~~ {\rm or}~~ r\la 0.15~\mathrm{fm},
\label{55}
 \ee
 In what
follows the region (\ref{55}) is called as the {\it transition region}. By
direct calculations of the integral (\ref{23}) it can be shown that
 at longer distances $\tilde \Lambda_R(r)$ is
approaching the vector constant $\Lambda_V$ \cite{14}:
\be
\tilde \Lambda_R\to \Lambda_V ~~ {\rm at}~~ r\ga 0.15~\mathrm{fm}.
\label{56}
 \ee

\section{Static interquark potential}

Knowing the differences
$\Delta \alpha_B^{(n)}(r)\equiv
\tilde\alpha_B^{(n)}(r)-\tilde\alpha_V^{(n)}(r)$ one can calculate the
corresponding differences  $\Delta V_B^{(n)}(r)$ between the  background and
perturbative static  potentials: \be
 \Delta V_B^{(n)}(r)=-\frac43\frac{\Delta\alpha_B^{(n)}(r)}{r},~~ (n=1,2).
\label{60a}
 \ee

 Numerically calculated $\Delta
V_B^{(1,2)}(r)$ are shown in Figs. 5,6 correspondingly.
We observe the linear rise of potentials illustrated by
the tangents (dashed lines). Tangent slopes $\sigma_B^{(n)}(r)$ are defined
as
\be
\sigma_B^{*(n)}(r)=\frac{\partial \Delta V_B^{(n)}(r)}{\partial r}.
\label{61}
\ee
We see from the figures that
 \be
\sigma_B^{*(1)}=1.20~\mathrm{GeV^2},~~0.05~\mathrm{fm}<r<0.12~\mathrm{fm},
\label{62a}
\ee
\be
\sigma_B^{*(2)}=0.87~\mathrm{GeV^2},~~0.03~\mathrm{fm}<r<0.09~\mathrm{fm}.
\label{63a}
\ee
 This large difference
means that the perturbative $\tilde\alpha_V(r)$ fails to describe lattice
data in this region. Nevertheless, we try here to compare our results
(\ref{62a}),(\ref{63a}) with the lattice ones from Ref.[18]. After
perturbative potential subtraction from lattice static quenched potential,
lattice data in the region 0.03 fm $<r<$0.15 fm were parameterized by linear
potential with the slope
\be
\sigma_{\mathrm{lat}}^*=(1.20\pm 0.36)~\mathrm{GeV}^2,
~~0.03~\mathrm{fm}<r<0.15~\mathrm{fm}.
\label{63}
\ee
One can see that $\sigma_{\mathrm{lat}}^*$ corresponds to
$\sigma_B^{*(1,2)}$ within one standard error. We conclude that this  large
linear slope of potential is well explained using $\alpha_B(r)$ instead
of $\tilde\alpha_V(r)$. In Fig. 7  we compare the lattice static potential
from Ref.[18] at the distances $0.05<r/r_0<0.45$ with the potential
\be
   V(r)=-\frac4 3 \frac{\alpha_B^{(2)}(r)}{r}+\sigma r+C,
\label{64}
\ee
where $\sigma=0.2$ GeV$^2$ and $C=-253$
MeV (shown by solid line). This potential includes background perturbative
potential $V_B^{(2)}(r)$ and linear confining potential $\sigma r$.
Constant $C$ corresponds to the quark self-energy.
 We observe that this potential describes all lattice data remarkably well.

At the
Fig.7 {\it 1-loop} and {\it 2-loop perturbative potentials}, calculated in
Ref.[18], are also shown. The difference between them and lattice points is
large and explained above. The {\it 1-loop perturbative + linear
potential} with the slope $\sigma_{\mathrm{lat}}^*$ is also shown. It
describes lattice points up to $r/r_0=0.35$, but fails to describe the rest
of data because of the Landau pole of $\tilde\alpha_V(r)$.
 On the contrary, $V(r)$ (Eq.(\ref{64})) not only describes all the
presented in Fig.7 lattice data in the region $0.05<r/r_0<0.3$, but also all
lattice data available up to $r=3r_0$  with a reasonable accuracy.

What is the physics of the linear part of $V(r)$? In the framework of BPTh
\cite{27} it was shown that the linear confining potential
starts from the quark distances close to the gluonic correlation length
$T_g$. From the lattice data $T_g\approx 0.2$ fm \cite{28} and
$T_g=0.12\div 0.15$ fm \cite{29}. At smaller distances one needs to take
into account an interference of perturbative and nonperturbative effects
\cite{30}. As was shown in Ref.\cite{30}, the interference potential at
$r\la T_g$ behaves like a linear one with the slope about $\sigma$, while
nonperturbative potential is proportional to $r^2$ and small in this region
\cite{27}. At distances $r\ga T_g$,  the interference interaction vanishes,
while the nonperturbative potential becomes the linear one with the same
slope $\sigma$. As a result the sum of these potentials may be close to a
linear one with the standard slope $\sigma$ over all distances.
 This linear behavior of nonperturbative potential at short distances is
 important for the fine structure fit in charmonium [10]. Note also that
the linear potential $ V^{(\mathrm{NP})}(r)$ is consistent with the
"short string" potential generated either by the topologically defined
point-like monopoles or infinitely thin P-vortices within the Abelian Higgs
model \cite{31}. The distinguishing feature of last potential is that it
does not change its slope at all distances.
We
leave the detailed numerical analysis of this effect to the subsequent
paper.

\section{Conclusions}

In our paper  the strong coupling $\tilde \alpha_B(r)$ in coordinate space,
deduced in BPTh, is investigated and the corresponding perturbative
potential is compared to short range lattice static potential from Ref.
[18]. The following prominent features of $\tilde\alpha_B(r)$ are observed:

1. The background coupling attains the asymptotics of standard perturbative
coupling $\tilde\alpha_V(r)$ only at very short distances, $r<0.04$ fm,
where the QCD constant $\Lambda_R=\Lambda_V\exp\gamma_E$ and $\Lambda_V$,
the QCD constant in the $V$-scheme is considered to be a well
established number in quenched approximation.

2. At larger $r$ a function $\tilde \Lambda_R(r)$ plays  the role of the QCD
constant which at short distances, $m_B r<\gamma_E$, is approximately given
by
\be
    \tilde \Lambda_R(r)\approx \Lambda_R\exp(-m_B r+\frac{m_B^2 r^2}{4}),
\ee
where $m_B=1.0$ GeV is the background mass fixed by fit to fine structure
splittings of $1P$ and $2P$ levels in bottomonium. In fact the condition

\be
     m_B r -\frac{1}{4} m_B^2 r^2\leq \gamma_E
\ee
defines the narrow transition region: 0.05 fm $\leq r \leq$ 0.15 fm, where
$\tilde \Lambda_R(r)$ decreases almost twice, from the value
$\Lambda_R$ at $r\approx 0.05$ fm to the number close to $\Lambda_V$ at
$r$=0.15 fm.

3. The static potential $V_B(r)$ in BPTh, defined through
$\tilde\alpha_B(r)$ in usual way, Eq. (36), is in a good agreement
with quenched lattice static potential measured at short distances [18].

4. The specific behavior of $\tilde\alpha_B(r)$ in the transition  region
produces a linear rise of the difference $\Delta
V(r)=V_B(r)-V_P(r)$ with the slope $\sigma^*\sim$ 1 GeV$^2$ in accord
to lattice data. Moreover, we have obtained a very good agreement with the
lattice data using the sum $V_B(r)+\sigma r$ with $\sigma=0.2$ GeV$^2$.

5. At distances $r\ga$ 0.2 fm the function $\tilde \Lambda_R(r)$
turns out to be almost constant; $\tilde \Lambda_R(r)\approx \Lambda_V$ over
all distances, so that at $r>0.3$ fm the coupling $\tilde\alpha_B(r)$ fast
approaches the freezing value $\tilde\alpha_B(\infty)$. This fact helps
to understand why the use of $\tilde\alpha_V(r)$=const in static potential
of quark model [3] appears to be a good approximation in hadron
spectroscopy.

6. The freezing value of the background coupling coincides in momentum and
coordinate space and this statement can be checked in different processes in
low energy QCD. Our estimate for $\alpha_{\mathrm{fr}}$ is 0.53-0.60.

The authors are grateful to Yu.A.Simonov for many fruitful discussions. This
work has been supported by RFFI grant 00-02-17836.

\clearpage
\begin{figure}[t]
 \epsfxsize=12.5cm
  \centering
  \epsfbox{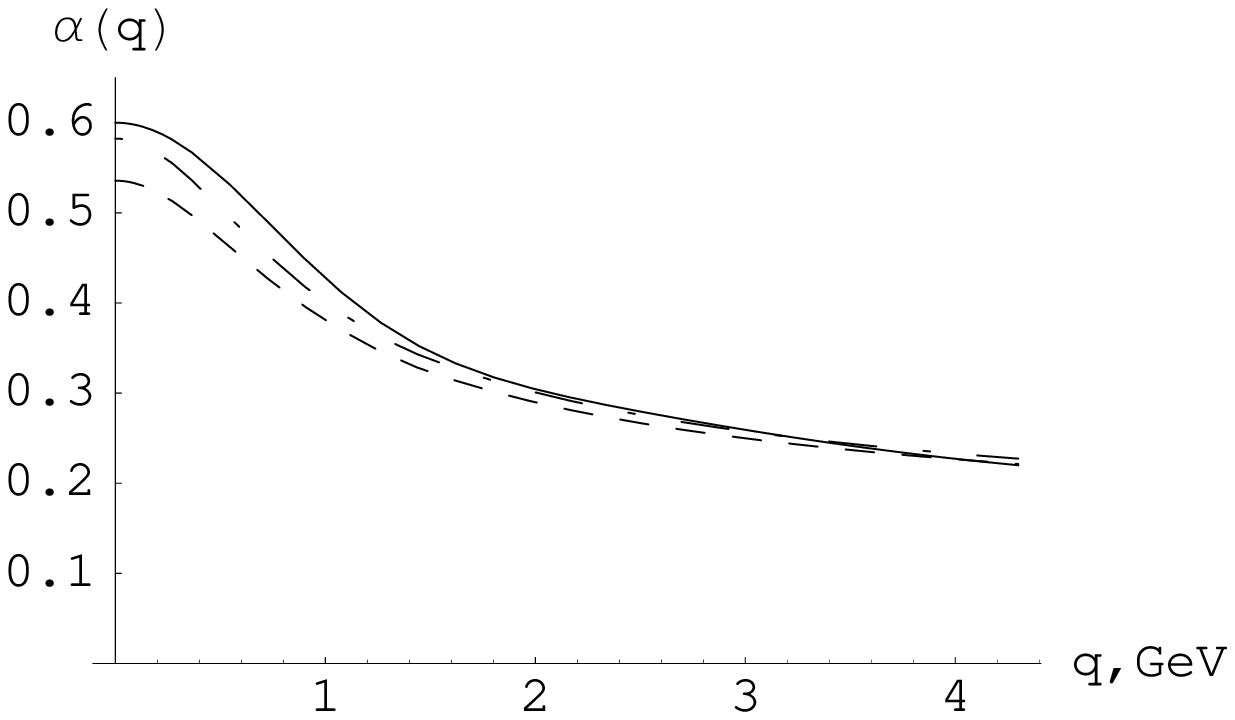}
   \caption{The behavior of the background coupling
$\alpha_B^{(2)}(q)$ for $\Lambda_V^{(3)}=385$ MeV (dashed line) and
  for $\Lambda_V^{(3)}=410$ MeV (dash-dotted line) compared to the
phenomenological coupling $\alpha_{\mathrm{GI}}(q)$ (solid line) taken from
Ref.[3]. } \end{figure}

  \begin{figure}[b]
 \epsfxsize=12.5cm
  \centering
  \epsfbox{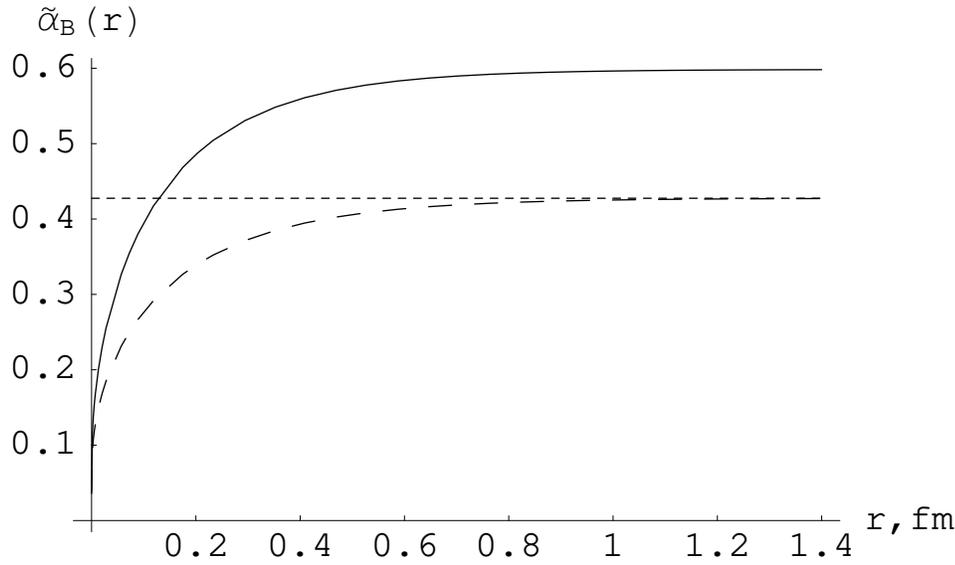}
   \caption{The 1-loop $\tilde \alpha_B^{(1)}(r)$ (solid line) and the
2-loop $\tilde \alpha_B^{(2)}(r)$ (dashed line) background couplings in
quenched approximation; in both cases $\Lambda_V^{(0)}=385$ MeV, $m_B=1.0$
GeV; 2-loop asymptotics, $\alpha_{\mathrm{cr}}^{(2)}=0.428$, is shown.}
 \end{figure}

\clearpage

 \begin{figure}[t]
 \epsfxsize=12.5cm
  \centering
  \epsfbox{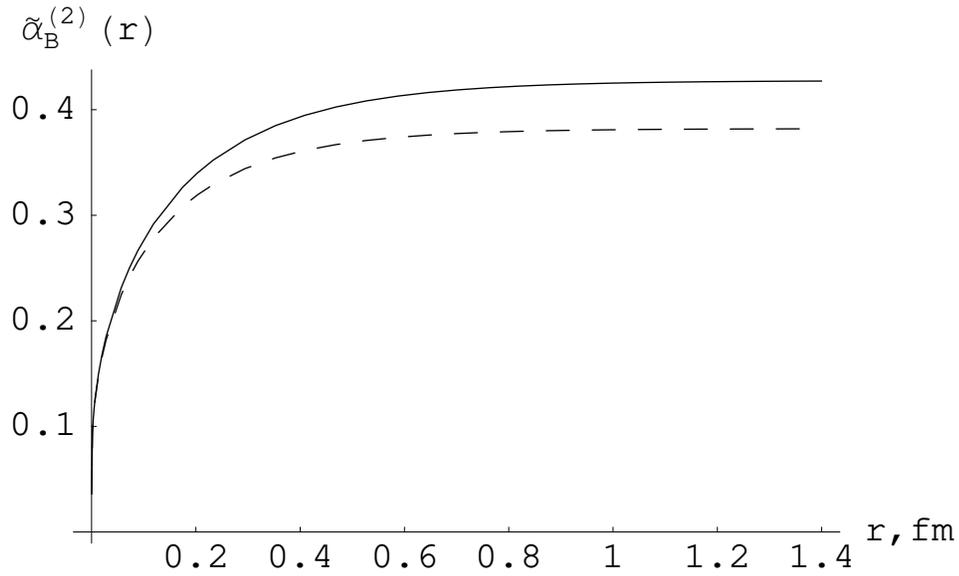}
   \caption{The background coupling $\tilde \alpha_B^{(2)}(r)$ in coordinate
space for two values of the background mass $m_B$: $m_B=1.0$ GeV (solid
line) and $m_B=1.1$ GeV (dashed line); in both cases  $\Lambda_V^{(0)}=385$
MeV.}
\end{figure}

 \begin{figure}[b]
  \epsfxsize=12.5cm
   \centering
   \epsfbox{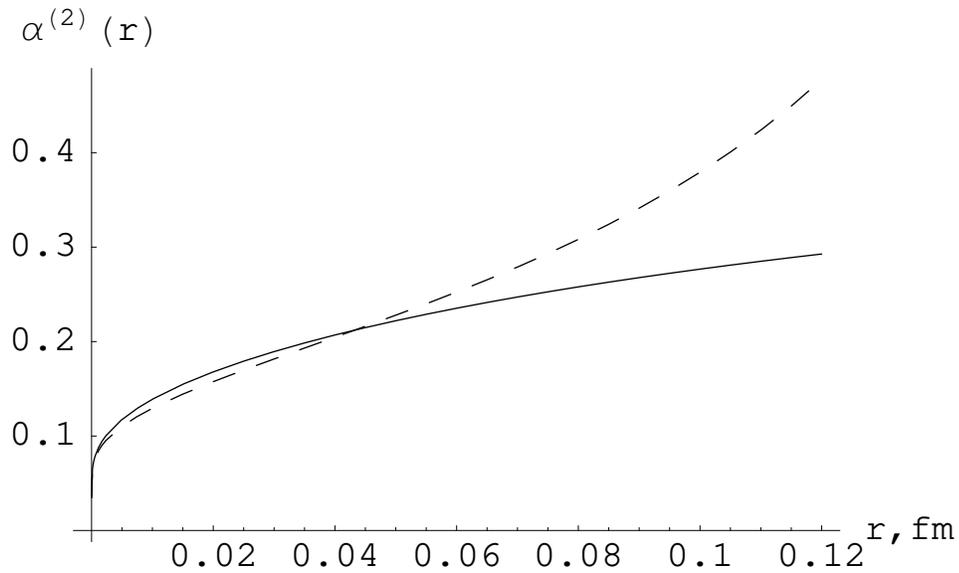}
   \caption{The  background coupling $\tilde \alpha_B^{(2)}(r)$ with
$\Lambda_V^{(0)}=385$ MeV, $m_B=1.0$ GeV (solid line)compared to the
perturbative  $\tilde\alpha_V^{(2)}(r)$ with $\Lambda_R=686$ MeV (dashed
line) at short distances.}
 \end{figure}

\clearpage

  \begin{figure}[t]
  \epsfxsize=12.5cm
   \centering
   \epsfbox{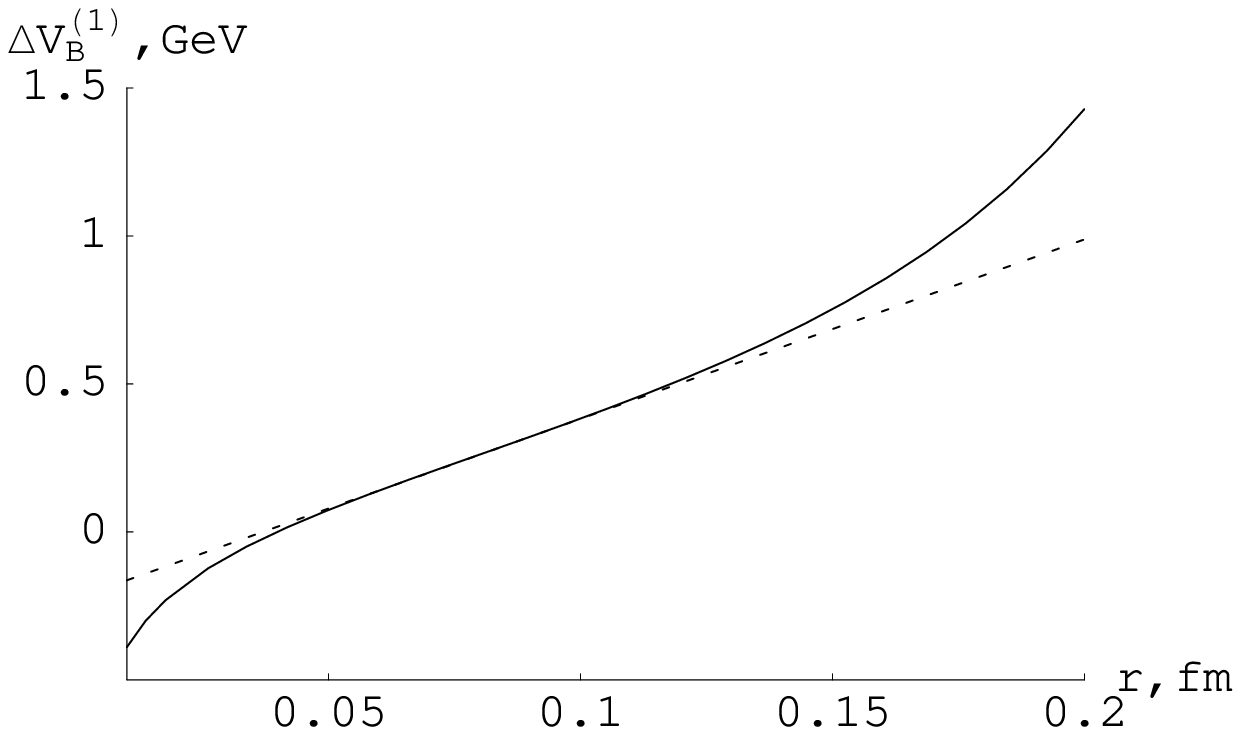}
   \caption{The difference $\Delta V_B^{(1)}$ between the background and
perturbative potentials in one-loop approximation (solid line). The tangent
with the slope $\sigma_B^{*(1)}=1.20$ GeV$^2$ is shown by the dashed line.
The parameteres are the same as in Fig. 4.}
 \end{figure}

  \begin{figure}[b]
  \epsfxsize=12.5cm
   \centering
   \epsfbox{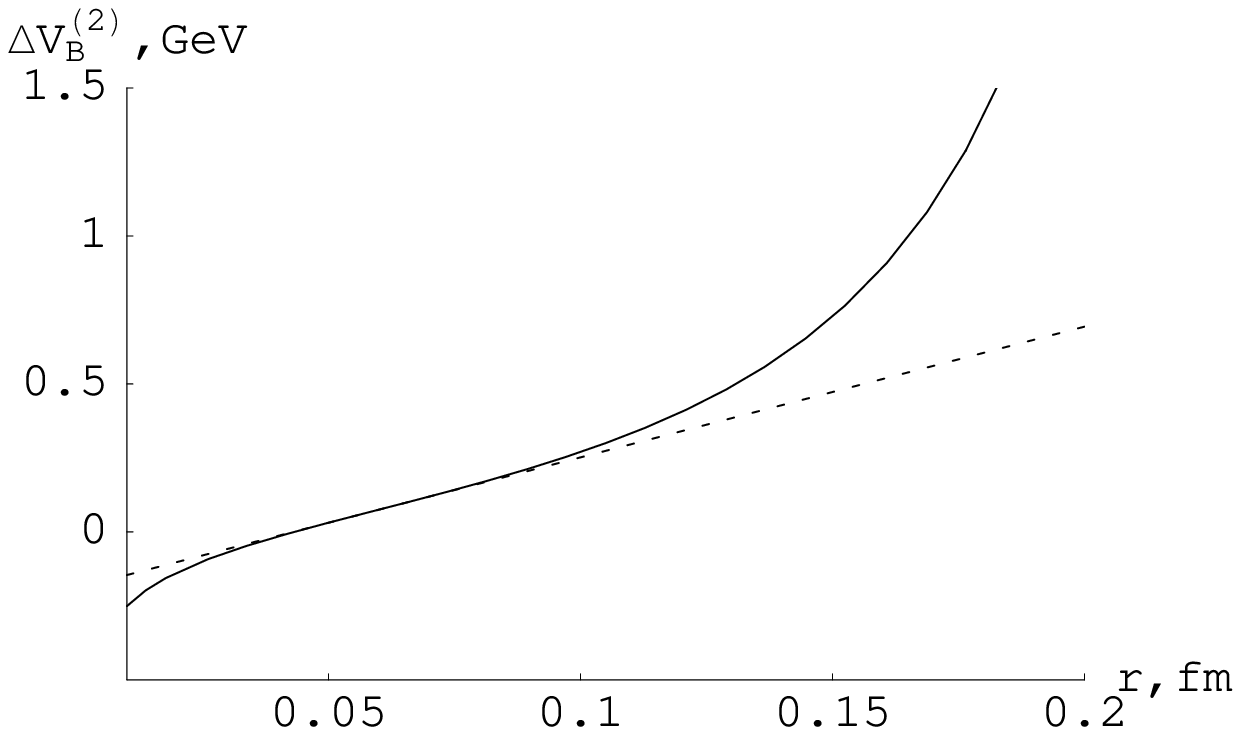}
   \caption{The difference $\Delta V_B^{(2)}$ between the background and
perturbative potentials in two-loop approximation (solid line). The tangent
with the slope $\sigma_B^{*(2)}=0.87$ GeV$^2$ is shown by the dashed line.
The parameteres are the same as in Fig. 4.}
   \end{figure}

 \end{document}